# Energy Bounds for Fault-Tolerant Nanoscale Designs

Diana Marculescu
Dept. of Electrical and Computer Engineering
Carnegie Mellon University
Pittsburgh, PA 15213

**Abstract** - *The problem of determining lower bounds for the energy cost of a given nanoscale design is addressed via a complexity theory-based approach. This paper provides a theoretical framework that is able to assess the trade-offs existing in nanoscale designs between the amount of redundancy needed for a given level of resilience to errors and the associated energy cost. Circuit size, logic depth and error resilience are analyzed and brought together in a theoretical framework that can be seamlessly integrated with automated synthesis tools and can guide the design process of nanoscale systems comprised of failure prone devices. The impact of redundancy addition on the switching energy and its relationship with leakage energy is modeled in detail. Results show that 99% error resilience is possible for fault-tolerant designs, but at the expense of at least 40% more energy if individual gates fail independently with probability of 1%.*

## 1. Introduction

Nanoscale computing for either CMOS or non-CMOS technology is widely agreed to be characterized by non-deterministic and unreliable behavior. In the case of nanometer CMOS designs, uncertainties stem from process and system parameter variability and their impact on performance and behavior of switching devices, as well as from increased susceptibility to soft errors which may thus create incorrect or faulty functionality. For non-CMOS nanoscale electronics (such as molecular or self-assembly), uncertainties in operation originate in the inherent non-deterministic switching behavior of such devices. While fault-tolerant computing is a mature area of research, automatically designing reliable systems out of unreliable components may prove to be a challenging task. As a natural solution to the fault-tolerance problem, the use of redundancy has been proposed and analyzed theoretically as early as five decades ago in the seminal work of von Neumann [1] and more recently by others [2,3,4,5,6,7]. While a practical solution to synthesizing in an automated manner nanoscale designs that are inherently fault-tolerant has yet to be found, nonetheless such theoretical results offer a good insight into the achievable limits of error-resilience and required minimum redundancy (or extra logic) needed. Furthermore, in the case of CMOS nanoscale technology, power density and energy cost have become the main design bottlenecks [8]. Adding redundancy in the hope of *increasing error-resilience* has the negative effect of eventually *decreasing the energy efficiency* of a given design. Thus, given the clear trade-off between achieving resilience via redundancy addition and energy efficiency, tools that can aid and guide the design process in a nanoscale design methodology become mandatory.

This paper is a step in this direction, by proposing *theoretic lower bounds* and trade-offs between energy efficiency and redundancy needed for achieving a certain error-resilience level. Such bounds and theoretical trade-offs provide valuable insights early in the design process and may aid and guide the automatic synthesis process of nanoscale systems. While the fault-tolerance aspects described in this paper are relevant for any type of nanoscale designs based on emerging technologies that inherently include non-determinism in normal operation, we will restrict ourselves to considering as a driver application computing systems that use electrons and energy barriers to represent or manipulate their states [9]. CMOS nanoscale designs are one such example. Based on first principle physics laws, it has been shown recently [9] that power density of such designs will reach in the next decade values close to 100 Watts/$cm^2$. These power densities, coupled with increased complexities and integration densities, put a lot more pressure on achieving more efficient designs in the nanoscale era. To address this problem, we consider the case of fault-tolerant nanoscale systems which need to rely on added redundancy to be able to cope with increased error rates. To characterize the available trade-offs between the required redundancy and the associated energy cost, we propose the use of a *complexity theory*-based framework.

## 2. Related Work and Paper Contribution

This paper addresses the impact of error-resilience and redundancy in a theoretical framework that can guide the design process for nanoscale systems. Our framework is able to provide bounds on the energy cost associated with achieving a given error-resilience level for a design based on error-prone gates (characterized by non-deterministic behavior), while also detailing when these bounds are likely to be tight. As opposed to the theoretical bounds developed by von Neumann and his successors [1,2,3,4,5,6], our approach does not assume any particular use of redundancy, such as majority voting or parallel restitution [1]. Instead, it is based on a framework relying on information theory concepts [7] to quantify the energy efficiency of nanoscale designs in the presence of errors.

While the problem of analyzing the fault-tolerance of nanoscale designs under given gate or wire error models has been recently started to gain interest [10], no such study has linked the error-resilience of a circuit to its energy efficiency. Recent work [11] has shown, using information theoretic arguments, that it is possible to trade-off energy efficiency and fault-tolerance through the use of voltage scaling. Indeed, by allowing the voltage to be reduced, power cost is reduced significantly, while noise level is increased, thereby allowing for trading-off energy and error-resilience. However, our goal is different. We propose to *quantify* how added redundancy necessary for achieving *reliable operation* out of unreliable components impacts overall *energy efficiency* of a nanoscale design. In addition, our approach is not bound to any particular type of redundancy (based on techniques such as majority voting or parallel restitution) and assumes that "noisy" gates and can be modeled as error-free devices cascaded with a *symmetric communication channel* modeling the error. In addition, while previous work [11] has linked energy efficiency and fault-tolerance in an information theoretic environment, it targets voltage scaling-reliability trade-offs and does not address entities that are more complex than a single gate or device. Our theoretical framework is able to completely characterize complex designs in terms of the energy-error-resilience trade-offs, while also providing insights into the relative contributions of static and switching energy.

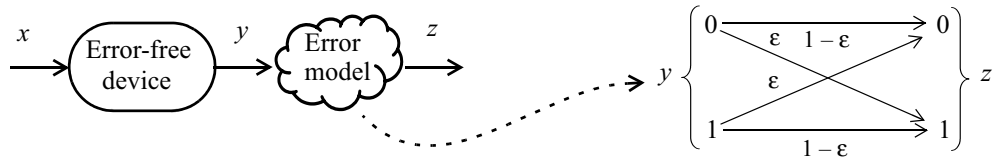

**Figure 1.** Modeling error-prone devices using error-free devices and a symmetric communication channel error model $\varepsilon$. The error-free device is a gate with a single or multiple bit input $x$ and a single bit output $y$.

## 3. Preliminaries

Without loss of generality, we restrict ourselves to nanoscale systems with the following characteristics:

- Circuits are assumed comprised of devices or gates that can be either error-free, or failure prone.[1]
- We consider the case of error-free and failure-prone circuits in which internal gates can fail independently with a probability $0 \leq \varepsilon \leq 1/2$.[2] We assume that any interconnect error probabilities are lumped into device error probabilities.
- We also consider device (gate) output error probabilities modeled as *symmetric communication channels* with an $\varepsilon$ error probability. As shown in Figure 1, each failure-prone device is assumed modeled as an error-free device cascaded with a symmetric communication channel.
- For the case of failure-prone circuits, we consider the case of $(1-\delta)$-*reliable computation*, that is, with probability of $1-\delta$, the output of the circuit is correct.
- While not restricting ourselves to any particular technology characterized by failure-prone operation, we consider the case of systems that use electrons and energy barriers to represent or manipulate their states. Examples include nanoscale CMOS applications or systems comprised of irreversible logic switch devices.

- Given this constraint, we are mostly concerned with *switching energy* or the energy lost due to information processing or transfer.

To this end, we consider the impact that any additional redundancy added to increase error resilience has on the overall energy efficiency of a nanoscale design. Assuming that most of the switching energy is consumed during output charging and discharging ($E = \frac{1}{2}CV_{dd}^2 sw$, where $E$ is the total energy consumed during a certain time interval, $V_{dd}$ is the supply voltage, $C$ is the load capacitance and $sw$ is the total switching activity during the considered interval), fault-tolerant circuits relying on added redundancy for increased error resilience can have *both* their load capacitance *and* switching activity affected. In addition, added redundancy may also impact logic depth, and thus overall latency. To this end, trade-offs between energy, delay or composite metrics (such as energy-delay product) on one hand, and required output error resilience $\delta$, on the other hand, can be uncovered and analyzed in an analytical manner.

## 4. Bounds on Switching Energy Metrics

Intuitively, added logic redundancy, not necessarily committed to any particular implementation (e.g., majority based voting used in triple or *n*-tuple modular redundancy vs. parallel restitution [1]) is expected to increase the error resilience of a failure-prone nanoscale design by providing multiple logical paths from the inputs to the outputs, thus decreasing the impact of the "noise" on the overall input-output communication channel. One may expect that the added redundant logic (when error-free) will perform in a similar manner as far as switching activity is concerned. However, we note that a non-zero error probability on the output of each device (be it part of the original circuit, or part of the redundant logic) can affect the switching behavior of that gate output, as shown next.

### 4.1. Impact on Switching Activity

In what follows, we denote by $p(x)$ the probability of a binary random variable $x$ being one and by $sw(x)$ the probability that $x$ changes state. We also denote by $\overline{x}$ an array of binary variables, and by $\overline{f}$ a multiple output Boolean function.

The following result quantifies the change in the switching activity that a given "noise" level produces on the output of a switching device:

*Theorem 1.* If $y, z$ are the error-free and error-prone outputs of a device which fails with a probability of $\varepsilon$ as in Figure 1, then:

$$sw(z) = (1-2\varepsilon)^2 sw(y) + 2\varepsilon(1-\varepsilon).$$

*Hint for proof:*[3] Uses switching activity definition for temporal

---

1. Devices and gates are assumed to denote the same entity in this paper.
2. As it will be seen in the sequel, if error probabilities are close to 1/2, reliable computation is not possible unless an unbounded overhead is considered.

3. Due to space limits, if not provided, all proofs can be found in [12].

independent signals: $sw(z) = 2p(z)(1-p(z))$. □

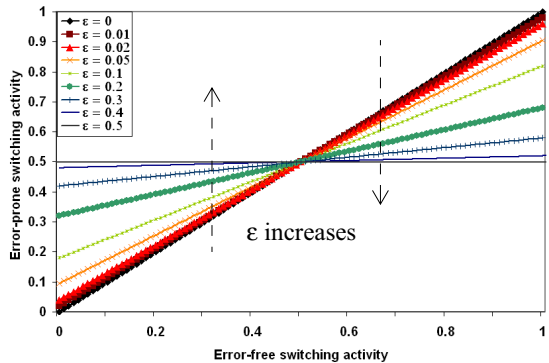

**Figure 2.** Switching activity of error-prone devices as a function of the switching activity of error-free devices.

It is interesting to note that failure-prone designs do not necessarily experience a monotonic trend as far as switching activity is concerned. As seen in Figure 2, while for smaller values of the error probability ε, the switching activity is less affected, for larger values (closer to 0.5) the switching activity becomes closer to 0.5, thus making gates with smaller switching activity more active and vice versa. In other words, increased error rates make the output of a failure-prone device look more random and less likely to carry any useful information.

### 4.2. Impact on Circuit Size

As shown in Theorem 1, any non-zero probability of error for a failure-prone device affects its average switching activity. In addition, added redundancy does not come for free: it affects both logic depth, as well as circuit size, which in turn affect performance and energy cost of the entire system.

In this section, we quantify the impact of added logic redundancy on circuit size, for a given error probability per gate ε, and a required maximum error probability on the output δ. The following result provides the tightest bounds known for the minimum circuit size required for computing a Boolean function with a given output fault-tolerance:

*Theorem 2.* [13] For $0 < \varepsilon \leq 1/2$ and $0 \leq \delta < 1/2$, if a Boolean function is $(1-\delta)$-reliably computed by a circuit with ε-noisy, $k$-input gates, then the additional redundancy in the circuit is at least:[1]

$$\frac{s \cdot \log s + 2s \cdot \log(2(1-2\delta))}{k \cdot \log t}$$

where $s$ is the sensitivity of Boolean function $f$ (or the maximum number of inputs that, if changed individually, produce a change on the output), $t = \frac{\omega^3 + (1-\omega)^3}{\omega(1-\omega)}$, and $\omega = \frac{1 - \sqrt[k]{1-2\varepsilon}}{2}$. □

The theorem confirms previous results [14,15] that show an $s \cdot \log s$ proportionality relation for the error-prone circuit size, but points to a superlinear increase in complexity with increased values of ε. Indeed, for ε values close to 0.5, circuit size approaches infinity, thus making $(1-\delta)$-reliable computation impossible for all practical purposes. On the other hand, it has been shown [2,3,5] that an upper bound on the size of a fault-tolerant circuit is $O(S_0 \cdot \log S_0)$ ($S_0$ is the size of the error-free implementation). For both the upper and lower bounds, equality is achieved for *parity functions*, implemented using decision trees or Shannon-like circuits. While the result above holds for single output circuits (Boolean functions), we extend it to general circuits below:

*Corollary 1.* For $0 < \varepsilon \leq 1/2$ and $0 \leq \delta < 1/2$, if an $m$-output function of $n$ inputs $\bar{f} : \{0,1\}^n \to \{0,1\}^m$ is $(1-\delta)$-reliably computed by a circuit with ε-noisy, $k$-input gates, then the additional redundancy in the circuit is at least:

$$\frac{s \cdot \log s + 2s \cdot \log(2(1-2\delta))}{k \cdot \log t}$$

where $s$ is the sensitivity of the multiple output function $\bar{f}$, and $t$, ω are as in Theorem 2.

*Hint for proof:* The characteristic function for $\bar{f}$ is a single output function with the same sensitivity. □

To understand the impact of increased gate count due to added redundancy, we show in Figure 3 the lower bound from Theorem 2 and Corollary 1 as a function of device error ε to $(1-\delta)$-reliably compute a 10-input parity function with sensitivity $s = 10$, error-free size $S_0 = 21$, and $\delta = 0.01$. We note that more than an order of magnitude redundancy factor is needed for error levels close to 0.5. It is thus expected that the increased size will also impact the total load capacitance and energy cost of the implementation.

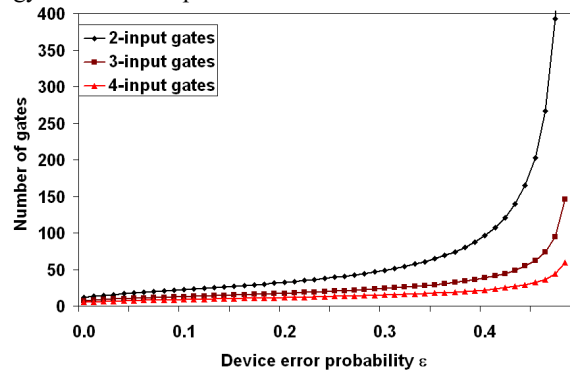

**Figure 3.** Minimum redundancy needed as a function of the device error ε, for $s = 10$, $S_0 = 21$, and $\delta = 0.01$ assuming 2-, 3-, and 4- input gate implementations.

### 4.3. Impact on Switching Energy

The impact of using redundancy on the energy of fault-tolerant nanoscale designs can be quantified through: 1) the change in the average switching activity for a generic internal gate; and 2) the change in the required number of devices needed to achieve a certain output reliability. Since we assume that, if considered error-free, redundant logic behaves almost the same as the original circuit in terms of switching activity, one can infer that Theorem 1 holds for the average switching activity of the error-free and error-prone circuits implementing the same function.

In addition, Corollary 1 can be used to quantify the increase in the total load capacitance of the fault-tolerant

---
1. In this paper, log is assumed base two, unless noted otherwise.

implementation: most high-level energy estimation frameworks have shown that total load capacitance is, with sufficient accuracy, proportional to the total device count [16,17]. Thus, one can conclude that the increase factor in energy for a fault-tolerant implementation satisfies the following:

*Corollary 2.* Consider $\overline{f} : \{0,1\}^n \rightarrow \{0,1\}^m$. The energy $E_{\varepsilon,\delta}$ of a circuit composed of ε-noisy, $k$-input gates, that $(1-\delta)$-reliably computes $\overline{f}$ satisfies:

$$\frac{E_{\varepsilon,\delta}}{E_0} \geq \left[1 + \frac{(\log s + 2 \cdot \log(2(1-2\delta)))}{k \cdot \log t} \cdot \frac{s}{S_0}\right] \cdot \left[(1-2\varepsilon)^2 + \frac{2\varepsilon(1-\varepsilon)}{sw_0}\right]$$

where $E_0$ is the energy cost of the error-free implementation, $s$ is the sensitivity of $\overline{f}$, $sw_0$ ($0 < sw_0 \leq 1$) is the average switching activity per gate of the error-free implementation, $S_0$ is the size of the error-free circuit implementing $\overline{f}$, and $t, \omega$ are as in Theorem 2.

*Hint for proof:* Uses Theorems 1-2 for energy estimation, assuming load capacitance is proportional to the total device count [16][17]. □

## 5. Case Study: Nanoscale CMOS Designs

The results presented in Section 4 hold true for the switching energy of any nanoscale computing system that relies on irreversible switching devices that use electrons and energy barriers to represent or manipulate their states. However, it is worth noting that, in the particular case of nanoscale CMOS designs (i.e., 0.09um and beyond) leakage power becomes the dominant component of the overall power, and thus increased redundancy implies increased number of idle devices or more leakage power. In addition, added logic redundancy comes at the expense of increased logic depth, which affects negatively circuit latency and overall performance. We address in the sequel some of these issues and the available trade-offs between switching energy bounds and leakage cost or performance.

### 5.1. Impact on Leakage Power

While the direct impact of logic redundancy on static power is clear through the increased number of potentially idle devices, there is a more subtle effect on leakage energy, due to changes in average switching activity per gate. More precisely, with probability $(1 - sw_{\varepsilon,\delta})$ a generic gate in the error-prone circuit is idle and subject to leakage energy, not switching energy. Thus, any change in the average switching activity triggers a change in the overall *contribution* of leakage energy to overall circuit energy. The next result details this change:

*Theorem 3.* If $W_{L,\varepsilon,\delta}$ is the ratio of leakage energy to switching energy in a circuit composed of ε-noisy gates, then:

$$\frac{W_{L,\varepsilon,\delta}}{W_{L,0}} = \frac{(1-2\varepsilon)^2 + 2\varepsilon(1-\varepsilon)/(1-sw_0)}{(1-2\varepsilon)^2 + 2\varepsilon(1-\varepsilon)/sw_0}$$

where $W_{L,0}$ is the ratio of leakage energy to switching energy in the error-free case and $sw_0$ ($0 < sw_0 < 1$) is the average switching activity per gate of the error-free implementation.

*Hint for proof:* Since switching energy $E_0 \propto S_0 V_{dd}^2 sw_0$ and static energy $E_{L,0} \propto (1 - sw_0) \cdot S_0 \cdot V_{dd} \cdot K$ ($S_0$ is the size of the circuit, $V_{dd}$ the supply voltage, and $K$ is a factor that depends on technology and threshold voltage), we can use Theorems 1-2 to determine switching/static energy in the error-prone case. □

To understand how the relative contribution to the overall energy cost changes, we show in Figure 4 the variation of the normalized ratio for the error-prone circuit, with respect to the error-free implementation for various switching activity values and error probabilities. As it can be noticed, the relative contribution of leakage energy *decreases* with increased error rates if the switching activity $sw_0$ is less than 0.5. Indeed, in this case, switching activity in the error-prone implementation increases and thus devices are less likely to be idle. The opposite is true if the switching activity of the error-free circuit exceeds 0.5. The relative contribution stays the same if $sw_0$ is exactly 0.5.

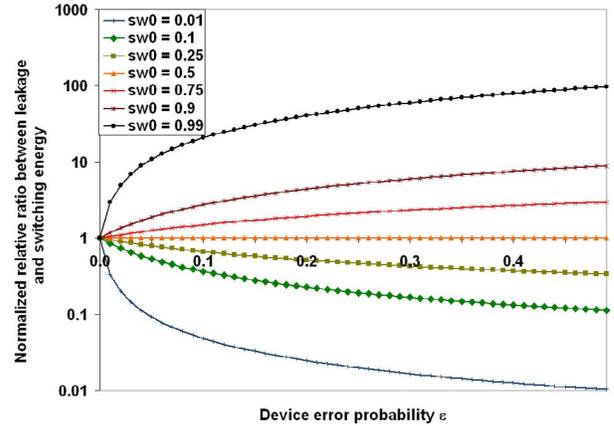

**Figure 4.** Normalized ratio between leakage energy and switching energy ($W_{L,\varepsilon,\delta}$) as a function of device error probability ε. The baseline is the ratio for the error-free implementation ($W_{L,0}$). Note the logarithmic scale on the Y-axis.

### 5.2. Impact on Performance, Average Power and Energy-Delay Product

Added redundancy is expected to have a significant effect on the circuit depth as well, and thus on overall latency. Indeed, the following result has been shown for circuit depth of fault-tolerant circuits:

*Theorem 4.* [7] Consider $\overline{f} : \{0,1\}^n \rightarrow \{0,1\}^m$. The logic depth $d_{\varepsilon,\delta}$ of a circuit composed of ε-noisy, $k$-input gates, that $(1-\delta)$-reliably computes $\overline{f}$ satisfies:

- If $\xi^2 > 1/k$, then $d_{\varepsilon,\delta} \geq \log(n\Delta)/\log(k\xi^2)$.
- If $\xi^2 \leq 1/k$, then $n \leq 1/\Delta$.

where $\varepsilon = (1-\xi)/2$ with $0 < \xi \leq 1$ [1] and $\Delta$ is defined as

$$\Delta = 1 + \delta\log\delta + (1-\delta)\log(1-\delta). \square$$

---
1. As shown in Section 4, only error levels that are less than 0.5 are of practical interest, thus the error probability is modeled as $(1-\xi)/2$.

In other words, to achieve a given error resilience (characterized by $\delta$), for sufficiently low error probabilities per gate, the minimum logic depth is inversely proportional to $\log(k\xi^2)$; otherwise, no circuit $(1-\delta)$-reliably computes $\overline{f}$, unless the number of inputs is less than $1/\Delta$. In the latter case, no known bounds are available for the size of the fault-tolerant implementation.

Analyzing these results, we note that, trade-offs between the *fault-tolerance* and *performance*, on one hand, and the average *power* consumption, or *energy×delay* product, on the other hand, can be deduced. Assuming that the latency of the failure-prone circuit varies as $D_{\varepsilon,\delta} \propto d_{\varepsilon,\delta} \cdot \frac{V_{dd}}{(V_{dd}-V_T)^\alpha}$ [18] (where $V_T$ is the threshold voltage and $\alpha$ is a technology dependent factor), if the same energy budget as the error-free circuit is targeted, the fault-tolerant implementation will need to rely on a lower $V_{dd}$ to compensate for the change in switched capacitance, which in turn further increases overall latency. Similar conclusions can be drawn if performance constraints need to be maintained instead: in this case, $V_{dd}$ must be increased to compensate for the logic depth increase, thus triggering an energy increase. As shown in Figure 5, the lower bound for normalized energy×delay product is higher than the one for normalized latency, for the same error level. This comes at no surprise, given the fact that both energy and delay increase for the fault-tolerant circuit implementation. However, the same cannot be said about the average power (or energy spent per unit time).

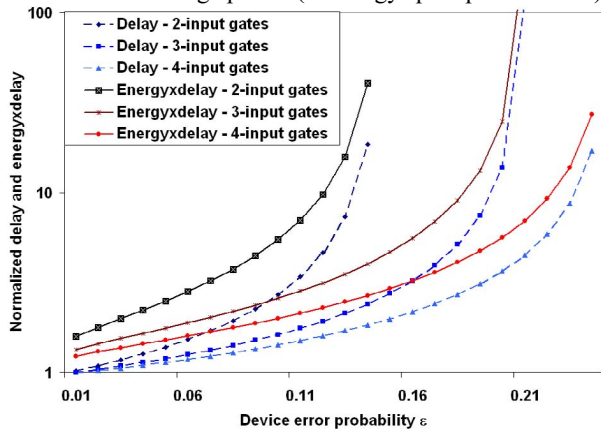

**Figure 5.** Normalized delay (energy×delay) as a function of $\varepsilon$. The baseline is the delay (energy×delay) for the error-free implementation. The contributions of the switching and leakage energy are assumed equal and switching activity is considered 0.5 in the baseline. All other parameters are as in Figure 3. Note the logarithmic scale on the Y-axis.

Figure 6 shows the dependency of the normalized average power on the device error probability and internal gates fanin. As it can be seen, for low error rates, the average power of the fault-tolerant implementation is larger than the one of the error-free circuit: in this case, delay increases at a lower rate than circuit size, and thus energy. Furthermore, a larger fanin reduces the overhead in average power. For larger error rates, however, the logic depth (and thus delay) increases at much higher rates than circuit size, thus making error-resilient implementations more power efficient, at the expense of significantly larger delays. In this case, larger fanins introduce a smaller power reduction as the difference in the increase rate for circuit depth and circuit size tapers off.

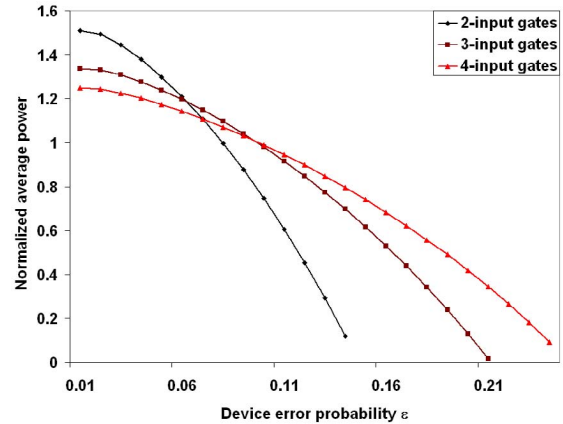

**Figure 6.** Normalized average power as a function of device error probability $\varepsilon$. The baseline is the average power for the error-free implementation. The contributions of the switching and leakage energy are assumed equal and the switching activity is considered 0.5 in the baseline. All other parameters are as in Figure 3.

## 6. Results

Although the theoretical results presented in previous sections are applicable *on average* to generic circuits, it is worthwhile to investigate how do these lower bounds apply for specific benchmarks. Even though all lower bounds presented are tight for a certain family of Boolean functions (mainly, parity functions implemented using decision-trees or based on a Shannon-like organization), when applied to specific circuits, these bounds still offer an idea about the possible trends that switching (and static) energy may have for fault-tolerant nanoscale designs.

To this end, we consider a subset of ISCAS'85 benchmarks and some computer arithmetic circuits (ripple-carry adders and array multipliers) with various bitwidths. The ISCAS'85 benchmarks have been optimized in the SIS [19] environment using *script.rugged.* All benchmarks have been mapped using a generic library comprised of gates with a maximum fanin of three. The average switching activity of a generic gate part of each benchmark has been obtained considering randomly generated inputs. For the error-free implementation we have assumed that 50% of the total energy is leakage (which is inline with predictions for technologies smaller than 0.09um [8]).

We show in Figure 7 the lower bounds for the energy and delay of the fault-tolerant implementation, normalized with respect to the error-free implementation. In each case, we consider three scenarios corresponding to gates independently failing with probability $\varepsilon = \{0.001, 0.01, 0.1\}$, and required maximum probability on the output $\delta = 0.01$. As it can be seen, the lower bounds increase significantly with higher error rates, necessitating in some cases at least 40% more energy if error rates are 1% per gate. We note that, while energy bounds are more circuit dependent (due to Boolean sensitivity *s* and switching activity $sw_0$), delay is less so, as the only circuit specific information it relies on is the average fanin *k*. Figure 8 shows the trend for the average power and energy×delay lower bounds, normalized with respect to the error-free implementation. While energy×delay lower bound experiences up to a 2.8X increase, average power is reduced due to the significant increase in logic depth (and latency).

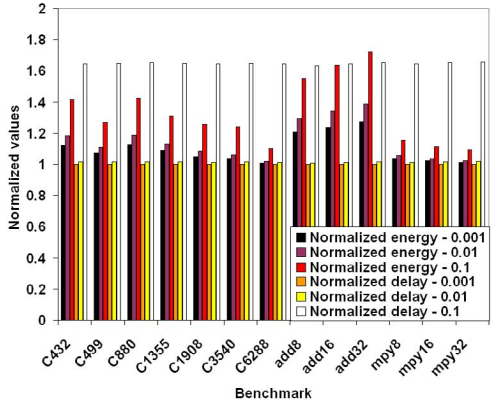

**Figure 7.** Lower bounds for energy and delay for ε = {0.001, 0.01, 0.1} and δ = 0.01, normalized with respect to the error-free implementation. The contributions of switching and leakage energy are assumed equal in the error-free baseline.

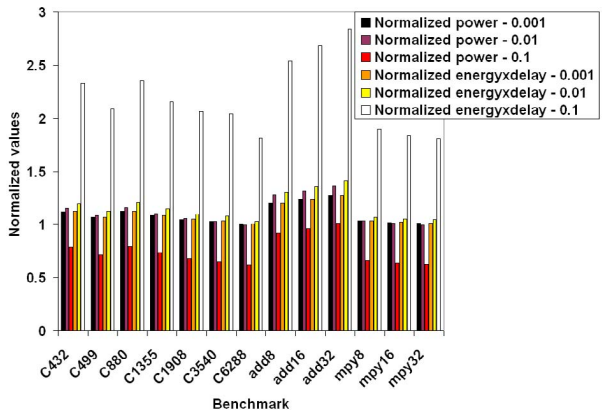

**Figure 8.** Lower bounds for average power and energy×delay for ε = {0.001, 0.01, 0.1} and δ = 0.01, normalized with respect to the error-free implementation. The contributions of switching and leakage energy are assumed equal in the error-free baseline.

## 7. Conclusion

This paper has introduced the use of complexity theory concepts for determining lower bounds on energy and related metrics for nanoscale, fault-tolerant designs. The theoretical results presented herein can be used to guide the automatic synthesis of error-prone nanoscale designs, while also uncovering the interplay between switching and static energy for these systems. Future work includes the treatment of sequential circuits and the refinement of the lower bounds depending on the circuit functionality.